\documentclass[preprintnumbers,amsmath,amssymb,twocolumn]{revtex4}
\usepackage[dvips,colorlinks=true,citecolor=red,filecolor=green,linkcolor=blue,pdfnewwindow=true]{hyperref}
\usepackage{graphicx}
\usepackage[title]{appendix}
\usepackage{color}

\begin{document}

\title{Scaling in topological properties of brain networks}
\author{Soibam Shyamchand Singh$^{1,4}$, Khundrakpam Budhachandra Singh$^{2}$, Romana Ishrat$^{1}$, B. Indrajit Sharma$^3$ and R.K. Brojen Singh$^{4}$}
\email{brojen@jnu.ac.in (Corresponding author)}
\affiliation{$^1$Centre for Interdisciplinary Research in Basic Sciences, Jamia Millia Islamia, New Delhi-110025, India.\\
$^2$McConnell Brain Imaging Center, Montreal Neurological Institute, McGill University, Montreal, QC H3A 2B4, Canada.\\
$^3$Department of Physics, Assam University, Silchar-788011, Assam, India.\\
$^4$School of Computational and Integrative Sciences, Jawaharlal Nehru University, New Delhi-110067, India.}

\begin{abstract}
The organization in brain networks shows highly modular features with weak inter-modular interaction. The topology of the networks involves emergence of modules and sub-modules at different levels of constitution governed by fractal laws. The modular organization, in terms of modular mass, inter-modular, and intra-modular interaction, also obeys fractal nature. The parameters which characterize topological properties of brain networks follow one parameter scaling theory in all levels of network structure which reveals the self-similar rules governing the network structure. The calculated fractal dimensions of brain networks of different species are found to decrease when one goes from lower to higher level species which implicates the more ordered and self-organized topography at higher level species. The sparsely distributed hubs in brain networks may be most influencing nodes but their absence may not cause network breakdown, and centrality parameters characterizing them also follow one parameter scaling law indicating self-similar roles of these hubs at different levels of organization in brain networks.
\end{abstract}


\maketitle

One of the most important issues in the study of brain networks is the origin of functional modules, organization of these modules and their functional relationships. Brain networks, constructed from various experimental studies on brains of different species, exhibit hierarchical features (highly modular structure) \cite{meu,bas}, and these modules are believed to be sufficiently isolated to enable them to perform independent functions \cite{ton}. These sparsely distributed modules in brain network are shown to exhibit small-world topology, which have large local clustering co-efficients and very small path lengths \cite{bas1}, and it may allow the modules to perform independent functions \cite{spor}. On the other hand, studies on brain networks derived from functional magnetic resonance imaging (fMRI) \cite{ueh,ach,ach1}, structural MRI \cite{hey} and diffusion tensor MRI \cite{hag} show small worldness in brain networks which seems inconsistent with the observed high modularity. High clustering in small worldness, which is a local parameter, could not explain the global high modularity of brain networks \cite{meu,for,gall}; and short path length in small worldness is also not suitable for strong modularity \cite{song,gall}. Since the strong modularity corresponds to \textit{large world}, the hierarchically organized, highly clustered, nearly isolated and self-similar set of modules are shown to be weakly tied among themselves \cite{song}, as a consequence of which the network preserves the small-world properties \cite{gall}. Therefore, the weak ties among the modules are believed to maximize the information transfer among the modules with minimum wiring cost, and also allow to maintain small-world topological characteristics \cite{gall,hilg}. Further, these weak connections among modules in brain networks compel limited propagation of avalanche of neurons among the modules and are modular size dependent \cite{russ}. 

Fractality or self-similar structures in a complex network could be one property which can explain functional relationships of a larger network down to the fundamental structure through different levels of organization \cite{song,song1}. Scaling and renormalization theory can probably highlight the importance of information flow in a complex network and its self-organization \cite{roz}. It has been shown that hierarchical organization of modules in functional brain networks (fMRI) show fractal properties \cite{gall}. This fractal organization of modules in the network keeps hubs tightly bound inside corresponding modules, and use low-degree nodes as inter-modular connectors showing \textit{disassortative topology} \cite{barab}. However, whether organization of modules and sub-modules at different topological levels follow fractal nature or not is still an open question. Further, whether the fractal properties exist in brain networks of different organisms (lower to higher level organisms) is not fully investigated.

In this study, brain networks of three different species (from lower to higher level of brain organization), namely, \textit{C. elegans}, cat, and macaque monkey were investigated. Scaling laws and fractal rules were applied on several topological parameters to investigate the self-organization and fractal properties of the brain networks. 
\begin{figure*}
\label{fig1}
\begin{center}
\includegraphics[height=17cm,width=11.0cm]{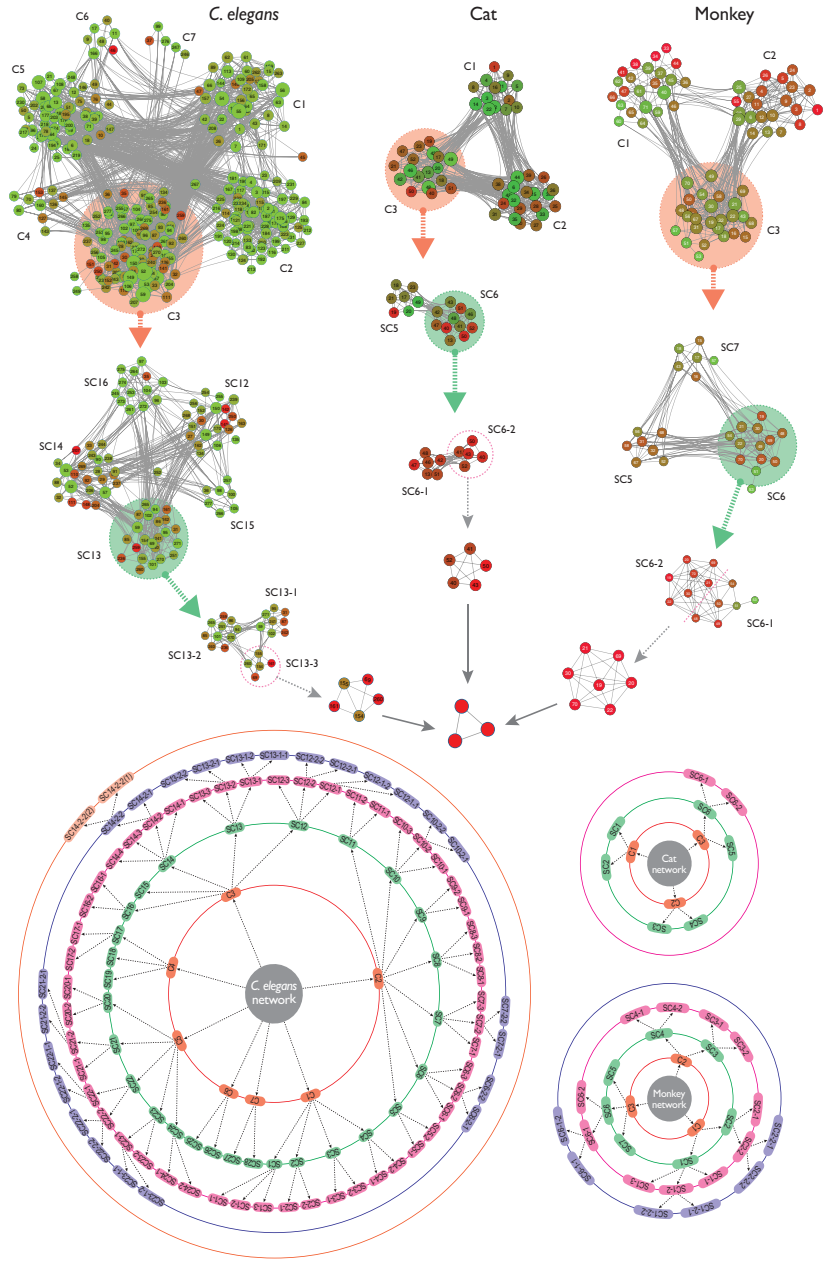}
\caption{Hierarchical organization of brain networks of \textit{C. elegans}, cat, and monkey at different levels. The upper parts show the topological arrangement of modules and sub-modules at various levels of organization (one way of largest module and sub-module) till the motif level. The lower parts show the organization of all modules and sub-modules at different levels (levels are indicated by circles) of the brain networks of the three species.} 
\end{center}
\end{figure*}

\vskip 0.3cm
{\noindent}\textbf{\large Results}\\
{\noindent}The topological properties of hierarchical network, which involve emergence of well-defined modules with few sparsely distributed hubs, can be characterized by three topological parameters, namely, probability of degree distribution $P(k)$, $P(k)\sim k^{-\gamma}$, with $\gamma\le 2.0$ \cite{rav}, clustering co-efficient $C(k)$, $C(k)\sim k^{-\alpha}$, with $\alpha\approx 1$ \cite{rav1}, and neighborhood connectivity $C_n(k)$, $C_n(k)\sim k^{-\beta}$ with $\beta\approx 0.5$ \cite{rom}; and follow power-law distributions with degree $k$ \cite{bara}. The set of the exponents $(\gamma,\alpha,\beta)$ of the three distributions $(P(k),~C(k),~C_n(k))$ calculated using network theory for the three species, namely, \textit{C. elegans}, cat, and monkey are given by: \textit{C. elegans} $\rightarrow (2.0,0.65,0.28)$; cat $\rightarrow (1.8,0.8,0.25)$; and monkey $\rightarrow (1.7,0.73,0.2)$, showing hierarchical features in all the three brain networks studied (Fig. 1 and Fig. 2 panels (A)). The organization of modules and smaller modules at different \textit{levels} (\textit{level-1} modules are the set of modules constructed from network, \textit{level-2} sub-modules are the set of all sub-modules constructed from \textit{level-1} modules, and so on) of the three species (Fig. 1) shows the hierarchical properties. The smallest module in each brain network (\textit{C. elegans}, cat, and monkey), from which all the three distributions ($P(k)$, $C(k),$ and $C_n(k)$) can be calculated,  is found to be triangle motif (Fig. 1). So one can think of triangular motif as the fundamental regulator of each brain network.
\begin{figure*}
\label{fig2}
\begin{center}
\includegraphics[height=18.0cm,width=14.0cm]{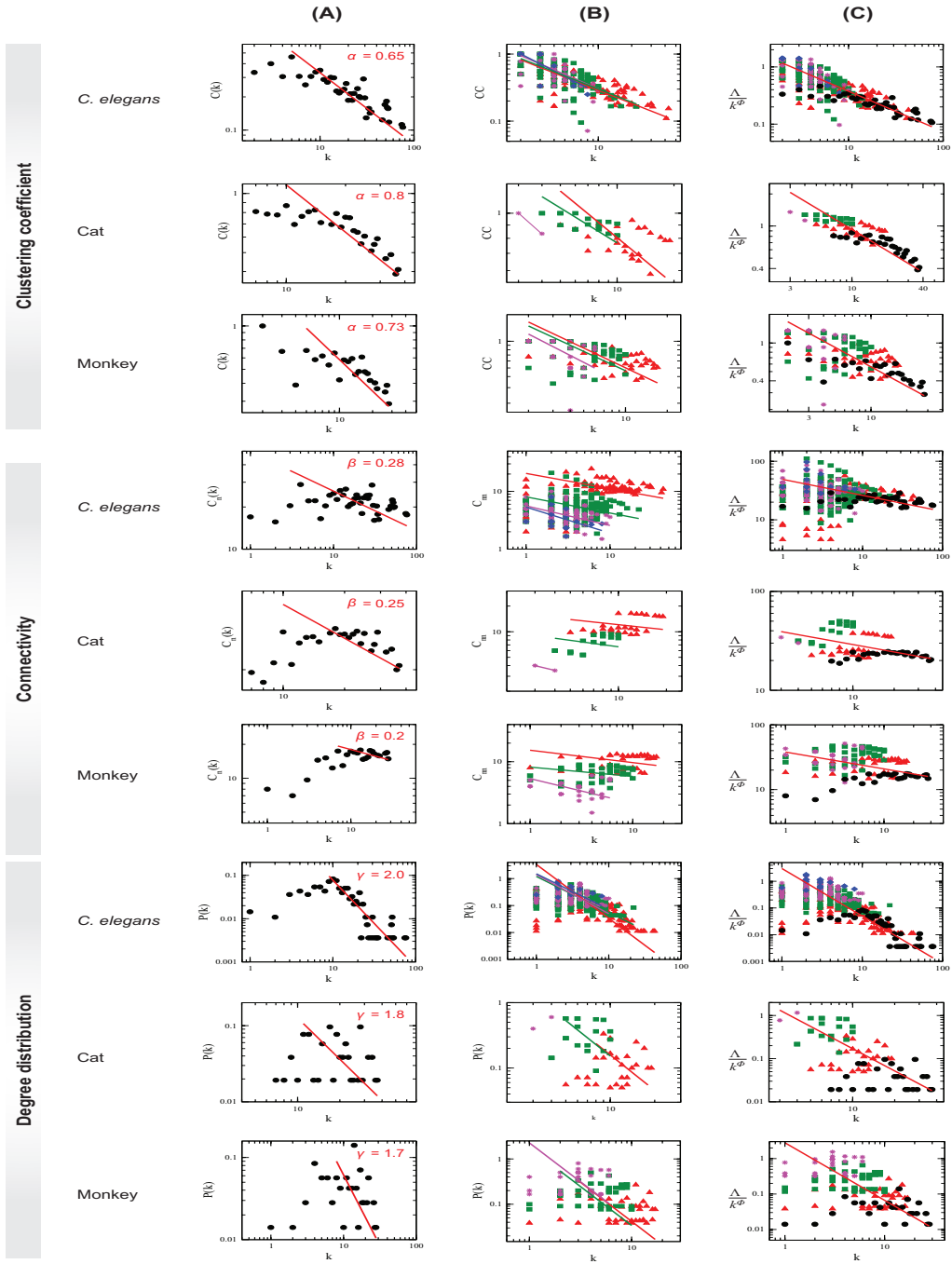}
\caption{Topological characteristics of brain networks of three species, \textit{C. elegans}, cat, and monkey: (A) for whole brain network, (B) for modules and sub-modules at various levels of network, (C) scaled for all modules and sub-modules in all levels to a single plot. The first three upper rows of panels are for clustering co-efficient, next three rows are for connectivity, and last three rows are for probability of degree distribution of the three species. } 
\end{center}
\end{figure*}

\vskip 0.3cm
{\noindent}\textbf{\bf Scaling of modules at different \textit{levels}}\\
We now study the topological properties of modules at different \textit{levels} in each brain network (Fig. 2 panels (A) and (B)). The plotted $P(k)$ versus $k$ of \textit{C. elegans} of larger modules and sub-modules at different \textit{levels} (Fig. 1: \textit{level-1}, C3; \textit{level-2}, SC13; \textit{level-3}, SC13-3) show nearly parallel straight lines of power-law fits of $P(k)$ with $k$ at log-log scale. The data points of the modules and sub-modules at different \textit{levels} along with data of whole network are rescaled to a single plot using one parameter scaling theory (see Methods) \cite{abra,pica,mack} and the power-law fitting of $P(k)$ on the rescaled data gives $\gamma=1.8$. This scaling behavior is satisfied to all the modules and sub-modules in the \textit{C. elegans} brain network, but shown only for one path as shown by arrows in Fig. 1. The same process is done to calculate clustering co-efficient $C(k)$ and neighborhood connectivity $C_n(k)$ of the same data set of modules and sub-modules at different \textit{levels}, rescaled the data set with original whole network using this one parameter scaling method, fitted with power-law equations with $k$ in log-log scale and found their exponents to be $\alpha=0.27$ and $\beta=0.7$, respectively. 

Similar network is done for cat and monkey brain networks also (Fig. 2). Scaling of the set of data of each species is done using the one parameter scaling method, and exponents of the respective distributions which specify topological characteristics of the networks are determined by fitting the rescaled data with the respective distribution equations (Fig. 2). The set of exponents of the distributions of the scaled data of the two species are: cat $\rightarrow (1.7,0.7,0.25)$ and monkey $\rightarrow (1.6,0.68,0.25),$ respectively.
\begin{figure}
\label{fig3}
\begin{center}
\includegraphics[height=10.0cm,width=7.0cm]{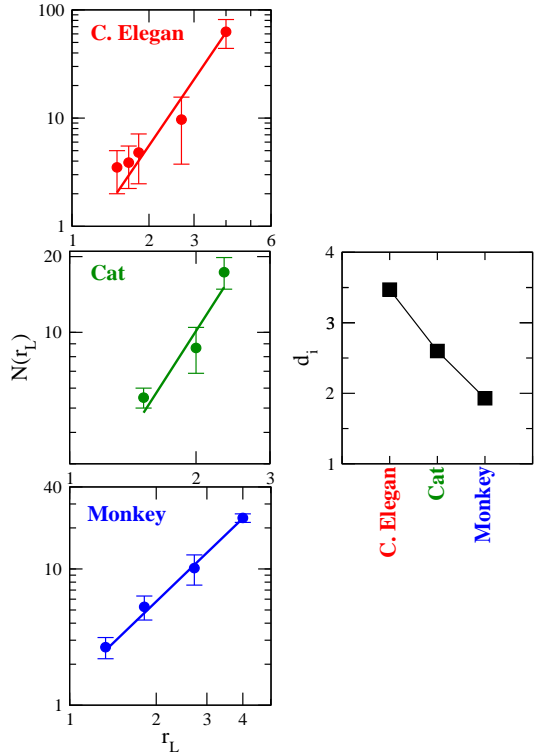}
\caption{Scaling behavior of modules and sub-modules at various levels of organization of the three species, \textit{C. elegans}, cat, and monkey, by calculating network mass (number of nodes) as a function of diameter. The right-hand panel is the fractal dimension of the three species.} 
\end{center}
\end{figure}

\vskip 0.3cm
{\noindent}\textbf{\bf Fractal nature of modules at different \textit{levels}}\\
The characterization of self-similar structures in network can be studied from the evolution of structures (number of nodes in the structures) in the network with path length \cite{song}. We calculated the number of nodes $n$ and diameter $R_L$ in each module or sub-module in a certain \textit{level L}, and then average over number of nodes $N=\langle n\rangle$ and path lengths $r_L=\langle R_L\rangle$ of all modules are taken. The evolution of $N(r_L)$ as a function of $r_L$ for all \textit{levels} in each network of the three species is shown in Fig. 3. The behavior of $N(r_L)$ with $r_L$ in all the brain networks follows the following power law:
\begin{eqnarray}
\label{frac}
N_i(r_L)\sim r_L^{d_i},
\end{eqnarray}
where $i=1,2,3$ and $d=\{d_i;i=1,2,3\}$ are Hausdorff fractal dimensions of brain networks of the three species. The value of $d_i$ of brain network of each species can be calculated by fitting the power-law equation (\ref{frac}) on $N(r_L)$ versus $r_L$ data of the respective species (Fig. 3), and the calculated fractal dimensions of all the three species are given in Fig. 3. The fractal dimension is found to be the largest for \textit{C. elegans} ($d_{ce}=3.47$) and smallest for monkey ($d_m=1.93$). Since fractal dimension is directly related to surface morphology of any system, larger value of fractal dimension may probably indicate larger disorder in network organization \cite{cor}. Its smaller value in the brain network of higher level species may reveal the organization of the network is more ordered and systematically self-organized \cite{seely}.  
\begin{figure}
\label{fig4}
\begin{center}
\includegraphics[height=10.0cm,width=7.0cm]{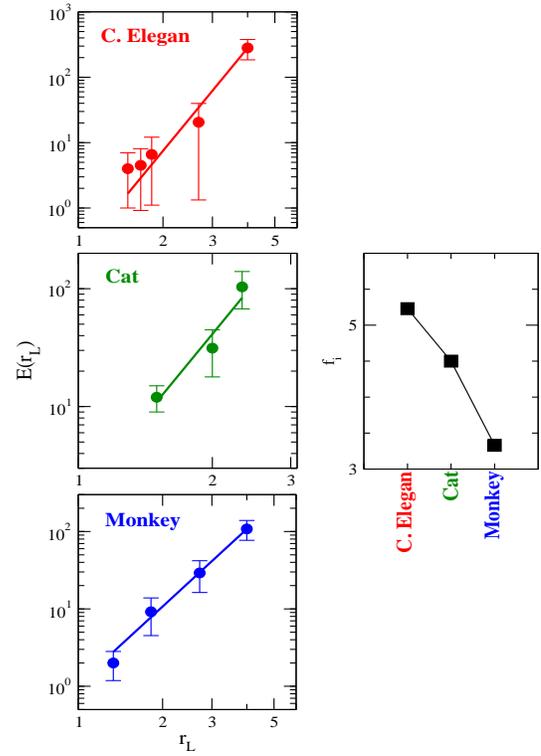}
\caption{Fractal nature of modules and sub-modules at various levels of organization of the three species, \textit{C. elegans}, cat, and monkey, by calculating the number of intra-edges as a function of diameter. The right-hand panel is the fractal dimension of the three species.} 
\end{center}
\end{figure}

To understand our claim of fractal nature of organization of modules (relating to the interaction) in brain networks, we now calculate the number of edges ($e$) and diameter ($r_L$) in each module of a certain \textit{level} (for $j$th module of $L$th \textit{level}: $e^{[j]}_L$, $r_L^{[j]}$), and then obtain average edges and diameter of the modules of the \textit{level} $L$ of $i$th species given by $E_i=\frac{1}{m}\sum_{j}^me^{[j]}_L$ and $r_L=\frac{1}{m}\sum_{j}^{m}r_L^{[j]}$, where $m$ is the number of modules/sub-modules at level $L$. The behavior of $E_i$ as a function of $r_L$ again obeys the following power law (Fig. 4):
\begin{eqnarray}
\label{edge}
E_i(r_L)\sim r_L^{f_i},
\end{eqnarray}
where $f_i=\{f_i;i=1,2,3\}$ is the set of fractal dimension relating to edges of modules and sub-modules of brain networks of the three species. The fractal dimension values of the respective species in this case are found to be higher than the respective values fractal dimension calculated using network mass or network node number, i.e. $f_i>d_i$; however, both $d_i$ and $f_i$ show the similar nature (Fig. 3 and Fig. 4 lower right panels).
\begin{figure}
\label{fig5}
\begin{center}
\includegraphics[height=10.0cm,width=7.0cm]{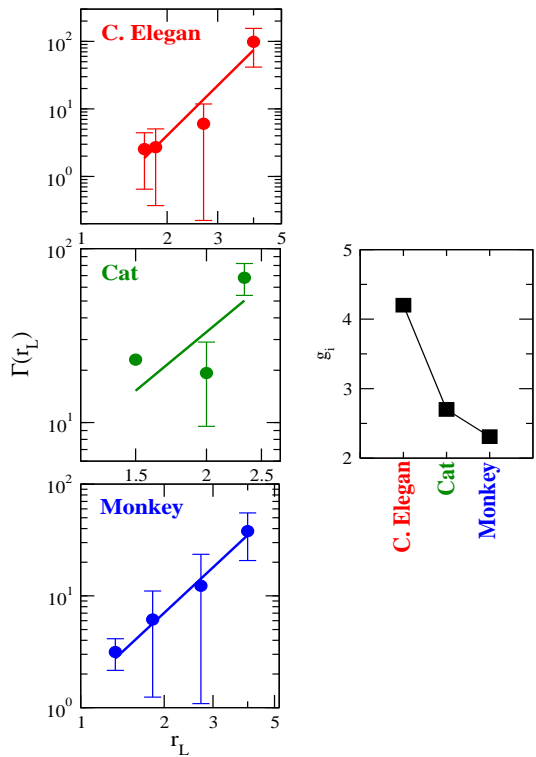}
\caption{Self-similar properties of modules and sub-modules at various levels of organization of the three species, \textit{C. elegans}, cat, and monkey, by calculating inter-modular edges of all modules and sub-modules as a function of diameter. The right-hand panel is the fractal dimension of the three species.} 
\end{center}
\end{figure}

The nature of organization of modules among different \textit{levels} can also be investigated by studying the inter-modular interaction among the modules and sub-modules. We calculate the number of edges between any pair of modules in a particular \textit{level} $L$ of brain network of $i$th species, average over all the inter-modular edges of all possible pairs of modules/sub-modules given by $\Gamma_i$, and then study the variation of $\Gamma_i$ as a function of average diameter of all modules/sub-modules $r_L$ in the level (Fig. 5). The variation of $\Gamma_i$ with $r_L$ for all brain networks of the three species \textit{C. elegans}, cat, and monkey (Fig. 5) shows power-law behavior (fitted line to the data points) given by
\begin{eqnarray}
\label{inter}
\Gamma_i(r_L)\sim r_L^{g_i},
\end{eqnarray}
where $g_i=\{g_i;i=1,2,3\}$ is the set of fractal dimensions of brain networks of all the three species. This power-law nature reveals the fractal nature of the inter-modular organization of the brain networks. The power-law behavior of mass (number of nodes), intra-modular and inter-modular edges of modules and sub-modules in all the \textit{levels} of brain networks show the fractal organization of brain networks.

The scaling and fractal properties of modules and sub-modules at different \textit{levels} of the brain network of each species probably connect the topological organization of the modules and sub-modules to their functionalities and working relationships among them, within and among the \textit{levels}. Further, self-organization among these modules and sub-modules could facilitate quick communication by minimizing the local and global energy expenditure in communication within the network. The increase in the value of fractal dimensions $d_i$ and $f_i$ given by equations (\ref{frac}) and (\ref{edge}) (Fig. 3 and Fig. 4) indicates the increase in complexity of the network \cite{cor,gall}. Since the values of $d_i$ and $f_i$ are minimum in monkey brain network as compare to the other two species, the modules and sub-modules in this network are more ordered and self-organized locally as well as globally as compared to the brain networks of the other two species. This efficient self-organization in the brain network of a certain species might reflect to the fast brain cognition in that species.
\begin{figure*}
\label{fig6}
\begin{center}
\includegraphics[height=18.0cm,width=14.0cm]{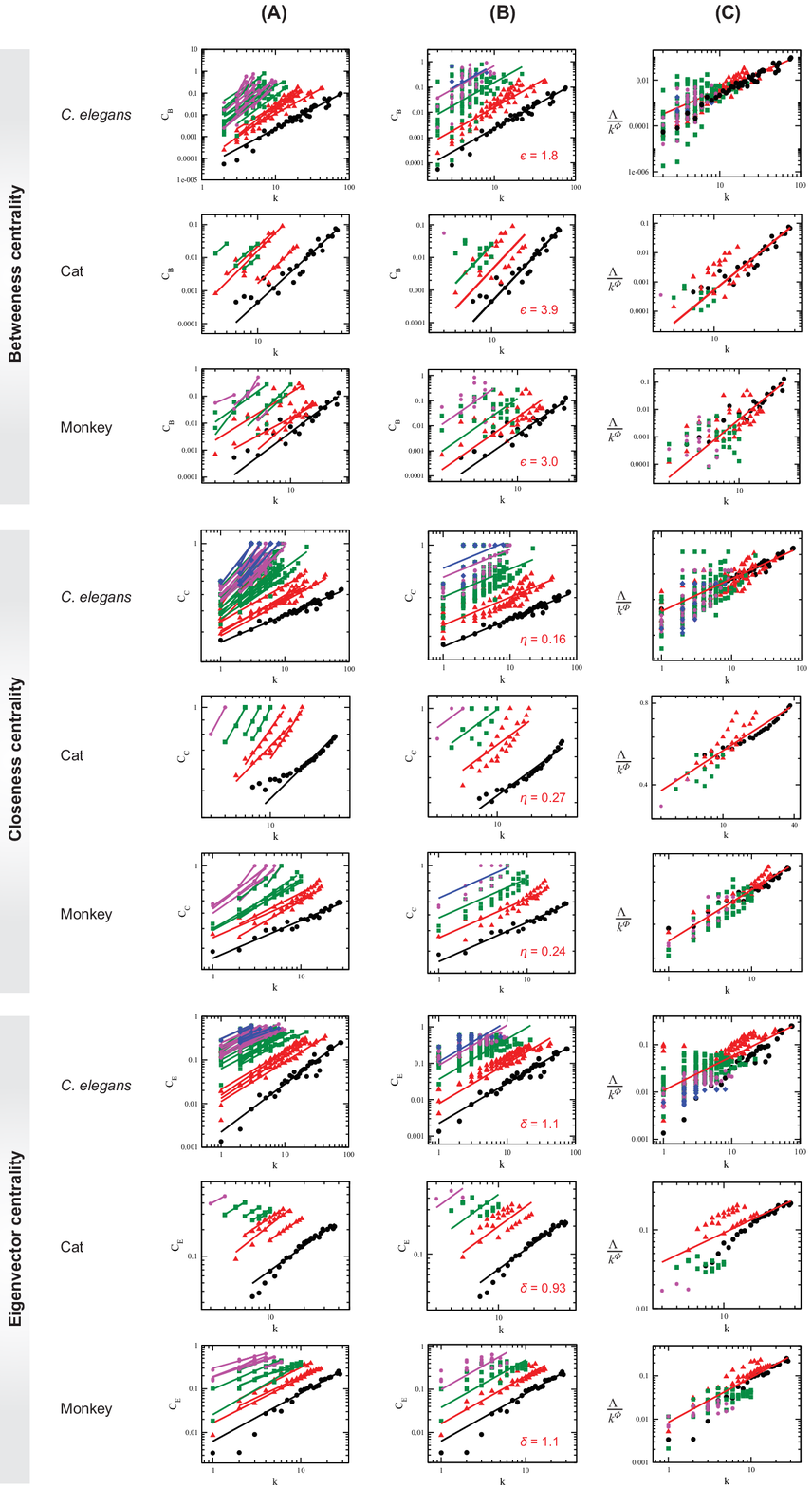}
\caption{Scaling in centrality parameters of brain networks of three species, \textit{C. elegans}, cat, and monkey: (A) centrality measures of all modules and sub-modules at various levels of the brain networks, (B) power-law fits on distribution of the centrality measures of each level, and (C) scaled centrality data of all modules and sub-modules into a single curve. The first three upper rows of panels are for betweenness centrality, next three rows are for closeness centrality, and last three rows are for eigenvalue centrality of the three species.} 
\end{center}
\end{figure*}

\vskip 0.3cm
{\noindent}\textbf{\bf Scaling in centralities and organization}\\
The betweenness centrality of \textit{C. elegans}, for the whole brain network, modules and sub-modules at different \textit{levels}, increases as degree of the network increases (Fig. 6 upper panel) which indicates that hubs in the network has significant roles in intra- and inter-modular/sub-modular signal processing at different \textit{levels}. Since high value of betweenness centrality of a node of degree $k$ reveals that the node could establish quick communication with other nodes in the network/module/sub-module through short paths \cite{ber,clar,bart}, hubs in the \textit{C. elegans} brain network may interfere in various network regulations and act as a controller of the network. Removing such few hubs emerged in the hierarchical brain may cause rewiring of the nodes in the modules and sub-modules at various \textit{levels} that may introduce new hierarchical topology of modules/sub-modules. The study of betweenness centrality of modules and sub-modules at different \textit{levels} $C_{\rm B}$ as a function of degree $k$ follows power-law distribution given by
\begin{eqnarray}
\label{bc}
C_{\rm B}(\epsilon_i)\sim k^{\epsilon_i},
\end{eqnarray}
where $\{\epsilon_i;i=1,2,3,4\}$ is the set of power-law exponents for different \textit{levels} indicated by $i$. The fitted lines on the data of modules and sub-modules show nearly parallel feature, and the values of $\epsilon_i$ are in the range $[1.23,2.12]$. It is also found that $C_{\rm B}$ of sub-modules increases significantly with smaller degree $k$ as the \textit{level} increases (Fig. 6 uppermost panel (A)). The data of all modules/sub-modules in a certain level are scaled using one parameter scaling method (see Methods section) to a single curve and fitted with equation (\ref{bc}), and it is found that all the four fitted lines on the four different \textit{levels} are approximately parallel (Fig. 6 uppermost panel (B)). Similarly, all the data of modules/sub-modules in all the levels are again scaled using the same method and fitted with equation (\ref{bc}) (Fig. 6 uppermost panel (C)), and the exponent is found to be $\epsilon=1.51$. This reveals that smaller modules have better communication among the nodes within each module, and hub/hubs in each module has similar roles (as fitted lines with power law on this smaller module is nearly parallel with the fitted lines on other modules/sub-modules of other \textit{levels}) with better performance (higher value of $C_{\rm B}$ at smaller value of $k$). However, as the sub-modules reach the motif level (here triangular motif), each node in the motif has equal importance (similar hubs due to same degree of each node), and therefore $C_{\rm B}$ of a motif, which is the smallest fundamental module, has a single value which is the largest as compared to that of other larger modules/sub-modules in the brain network. We did the same process of analysis to cat and monkey brain network data, and found similar behavior in $C_{\rm B}$ as a function of $k$ given by the scaling law of equation (\ref{bc}) (Fig. 6, second and third row panels (A), (B), and (C)), and their exponents are found to be $\epsilon=3.75$ and $\epsilon=3.02$, respectively. The scaling in the power-law behavior of $C_{\rm B}$ indicates fractal behavior of the modules/sub-modules at various levels up to the motif level.

Closeness centrality ($C_{\rm C}$) is another measure of centrality which describes how quickly an information from (by) a node can be propagated (received) to (from) the rest of the network, and can be characterized by the inverse of average distance between a given node with other nodes in the network \cite{newman2012}. The calculated $C_{\rm C}$ of \textit{C. elegans} as a function of degree $k$ increases as $k$ increases (Fig. 6, fourth row panel (A)) which indicates that the increase in $C_{\rm C}$ with $k$ exhibits shorter average path length (see equation (\ref{method_eq3}) in Methods) meaning faster information processing of the node with the rest of the brain network. This means that larger hubs (larger $k$) are able to communicate with the rest of the nodes in the brain network of \textit{C. elegans} faster than the nodes with smaller $k$, which is true for hubs in modules/sub-modules other than motif where every constituting nodes have same $k$. The data of modules and sub-modules at various levels obey the following power-law behavior:
\begin{eqnarray}
\label{cc}
C_{\rm C}(\eta_i)\sim k^{\eta_i},
\end{eqnarray}
where the set $\{\eta_i;i=1,2,3,4\}$ are closeness centrality exponents at various levels. The fitted curves with equation (\ref{cc}) on the data of modules and sub-modules at various levels of the network are approximately parallel (Fig. 6, fourth row panel (A)). Scaling of modules/sub-modules at each level are done using one parameter scaling method and then fitted with equation (\ref{cc}) (Fig. 6, fourth row panel (B)), and it is found that the fitted lines are approximately parallel. These scaled data of all levels are finally scaled with the same scaling method to a single curve and the scaled data obey the power law given by equation (\ref{cc}) (Fig. 6, fourth row panel (B)), and found $\eta=0.16$. The same procedure has been used to analyze the data of brain network of cat as well as monkey. The similar behavior in terms of scaling and structural properties are found in both the two species, and the two brain data follow the power-law behavior given by equation (\ref{cc}) with power-law exponent $\eta=0.28$ and $\eta=0.26$, respectively. The results show the fractal behavior of closeness centrality which may connect the network topology of brain networks to brain functionality.

Eigenvalue centrality (EC) is in favor of highly correlated nodes (which are usually high degree nodes) with rest of nodes in a network, and specific nodes which connect central nodes within the network relating to global network pattern \cite{lohm}. EC is characterized by well-connectedness in a network \cite{bona1}, a smooth enough function \cite{canr1}, and is a good measure of spreading (receiving) power of information of nodes in (from) the network \cite{canr2}. The calculated EC of the brain network of \textit{C. elegans} $(C_{\rm E})$ for the network, modules, and sub-modules at various levels (see Methods) show increase in its values as degree $k$ increases, obeying the following power law,
\begin{eqnarray}
\label{ce}
C_{\rm E}(\delta_i)\sim k^{\delta_i},
\end{eqnarray}
where $\{\delta_i;i=1,2,3,4\}$ is the set of EC exponents at various levels. As found in betweenness and closeness centralities, the fitted lines on the data of \textit{C. elegans} brain network, its modules, and sub-modules are nearly parallel with EC exponents in the range [0.5, 1.1]. Similarly, it is also found that as one goes towards higher levels, i.e. smaller module levels, $C_{\rm C}$ also increases comparatively. We then scaled the data of modules/sub-modules at each level, fitted with equation (\ref{ce}), and found that the fitted lines on the scaled data are also approximately parallel (Fig. 6, seventh row panel (B)) with $\delta\sim 1.1$. We then rescaled the data of modules and sub-modules in a single one and fitted with equation (\ref{ce}) (Fig. 6, seventh row panel (C)) and found $\delta\sim 0.0.72$.

Similar behavior is found in the brain networks of cat (Fig. 6, eighth row panels (A), (B), and (C)) and monkey (Fig. 6, ninth row panels (A), (B), and (C)) following the same scaling power law given by equation (\ref{ce}), and found the values of $\delta$ to be 0.71 and 1.03, respectively. 

\vskip 0.3cm
{\noindent}\textbf{\large Discussion}\\
The findings of our study suggest that the fundamental working principle of brain (in both lower and higher level species) is a system level topological self-organization. The fractal nature and scaling properties of these brain networks show self-similar organization of various topographical modules/sub-modules at every levels of constitution, which may relate to the functional brain organization, and energy cost in information transfer within and among the levels of organization is minimized. In addition, the few sparsely distributed hubs are tightly bound in their respective module and interfere functionalities of their own module, but could not influence rest of the modules at various levels in brain networks. In terms of inter-modular and intra-modular interaction edges, each brain network still show fractal nature which indicates systematic self-similar information processing at every levels and their interference. The decrease in the values of fractal dimension in going from \textit{C. elegans} to monkey (lower to higher species) shows that the organization of brain networks (in terms of signal processing, topological characteristics, and modular organization) is more ordered and self-organized systematically in higher species. Such topological properties in brain networks allow efficient information processing, constitution of fractal laws in the organization, and controlled behavior of hubs in the global network properties. 

The centrality measures (betweenness, closeness, and eigenvalue centralities) of brain networks, its modules, and sub-modules show increase in their values with degree showing that hubs behave as most influencing nodes in the modules/sub-modules they are embedded. These hubs act as central in the local module/sub-module, and they become local quick information spreader and receiver in the network. However, removing one hub in such situation does not cause the network breakdown because of the system level organization of the network through modules and smaller modules which are compact with their own fundamental rules. The centrality data of the brain network, modules, and sub-modules at different levels can be scaled into a single power-law behavior showing fractal nature. This exhibited fractal nature in the brain network could be the consequence of the emergence of a few most influencing hubs in each module/sub-module at any level of the network except at the level of motif where all the nodes in it have equal degree. Therefore, in the brain network, modules, and smaller modules, most popular node/nodes always exist and they take maximum responsibility in regulating the network/module/sub-module. However, these hubs' interference in the network is controlled (due to limited number of links to the modules/sub-modules) in such a way that they cannot control the other modules/sub-modules but can regulate them. 

The scaling properties in brain networks reveal complicated self-organization of the network at various topological levels, and it could probably explain systematic organization of functional modules via weak interaction among them. This topographic organization may induce the origin of brain functionalities even at the absence of few hubs or modules. However, the properties of this static network do not fully explain the working principle of the complicated brain network, its dynamics, and functional relationships. The studies on dynamics and multi-scaled network approach may highlight further interesting insights on brain organization/reorganization.

\vskip 0.3cm
{\noindent}\textbf{\Large Methods}\\
\vskip 0.05cm
{\noindent}\textbf{Data sources}\\
{\noindent}In this paper, the connection matrices of (1) \textit{C. elegans} neuronal system, (2) 52 cortical areas in cat species, and (3) 71 cortical areas in Macaque monkey species are studied.

The \textit{C. elegans} neuronal connectivity data set is adapted from Achacoso \& Yamamoto \cite{achacoso1992}, the compilation of which is based on the work of White et al. \cite{white1986} in which the neuronal connection were traced with electron microscope reconstructions. Further modifications are the removal of 20 neurons in the pharyngeal nervous system which have no internal connection information \cite{achacoso1992} and the additional removal of three other neurons (AIBL, AIYL, and SMDVL) considering their lack of spatial information \cite{kaiser2006,choe2004}. Finally, 277 neurons sharing 2102 synaptic connections are considered for further topological analysis (data set available at http://www.biological-networks.org).

The cat connection matrix used in this study is derived from the original article by Scannell et al. \cite{scannell1999}. In their paper, they collected information on the thalamo-cortico-cortical connections from many published studies, and applied the methods of non-metric multi-dimensional scaling, optimal set analysis, and non-parametric cluster analysis to derive the connection matrix of the 53 cortical areas and 42 thalamic nuclei. Their connection matrix is relatively weighted (0, 1, 2, and 3) according to the connection strength (absent/unreported, weak, intermediate, and strong, respectively) between each region. In this paper, only those connections among the 52 cortico-cortical areas are studied (after `Hipp' area is omitted)\cite{sporns2004}. The relative weighting is discarded and only the presence or absence of connection is considered in the respective adjacency matrix. The resulting final matrix has 52 cortical areas and 820 cortico-cortical connections. 

Collecting information from the neuroanatomical studies, Young \cite{young1993} applied the method of optimization analysis to map the cortico-cortical connections between 73 cortical areas of interest in the entire cerebral cortex of Macaque monkey. The connection matrix of Macaque monkey used in this paper is also based on the study of Young \cite{young1993}, with a modification as mentioned in Sporns \& Zwi \cite{sporns2004} in which two areas of interest (Hipp and Amyg) are removed resulting to a total of 71 cortical areas with 746 interconnections (data set available at https://sites.google.com/site/bctnet/datasets).

\vskip 0.3cm
{\noindent}\textbf{Graph construction and network parameters}\\
{\noindent}The connection matrices (adjacency matrices) from the above-mentioned data sets are used to generate  undirected graphs by using \textit{igraph} R package \cite{csardi2006}. For identifying communities in these graphs, the leading eigenvector spectral graph partitioning method (for which algorithm is available in \textit{igraph} package) is implemented \cite{newman2006}. In this method, the modularity term is expressed in terms of eigenvalues and eigenvectors of a modularity matrix, and the partitioning is done using multiple leading eigenvectors that optimizes the modularity \cite{chen2014}.  The communities are then grouped into each topological level. For each graph/subgraph we use the NetworkAnalyzer \cite{assenov2008,doncheva2012} and CytoNCA \cite{tang2015} plug-ins in \textit{Cytoscape}  for finding required network parameters such as degree, clustering coefficient, neighborhood connectivity, betweenness  centrality, closeness centrality, and eigenvector centrality.

\vskip 0.3cm
{\noindent}\textbf{\it 1. Degree distribution}\\
{\noindent}The degree represents a centrality measure that indicates the number of communications a node maintains with other nodes in a graph. The  degree distribution ($P(k)$) which is the probability that a randomly chosen node has a degree $k$ represents an important parameter that  helps us to identify whether a graph is random, scale free, hierarchical, etc. 
\begin{equation}
P(k) = \frac{n_k}{N},
\label{method_eq1}
\end{equation}
where $n_k$ represents the number of nodes with degree $k$ and $N$ is the total number of nodes in the graph.

\vskip 0.3cm
{\noindent}\textbf{\it 2. Neighborhood connectivity}\\
{\noindent}Neighborhood connectivity of a node $i$ represents the average connectivities (average degrees) of the nearest neighbors of node $i$ \cite{maslov2002}.  

\vskip 0.3cm
{\noindent}\textbf{\it 3. Clustering co-efficient}\\
{\noindent}Clustering co-efficient is a measure of  how strongly a node's neighborhoods are interconnected. Graph theoretically clustering coefficient is the ratio of the number of triangular motifs a node has with its nearest neighbor to the maximum possible number of such motifs. For an undirected graph, clustering coefficient ($C_i$) of the $i$th node can mathematically be expressed as
\begin{equation}
C_i = \frac{2e_i}{k_i (k_i -1)},
\label{method_eq2}
\end{equation} 
where $e_i$ is the number of connected pairs of nearest-neighbor of the $i$th node, and $k_i$ is the degree of the $i$th node.

\vskip 0.3cm
{\noindent}\textbf{\it Centrality measurement}\\
{\noindent}In addition, other important centrality measures include (1) closeness centrality, (2) betweenness centrality, and (3) eigenvector centrality. Centrality measures are helpful in identifying influential node(s) in a graph. 

\vskip 0.3cm
{\noindent}\textbf{\it 4. Closeness centrality}\\
{\noindent}Closeness centrality ($C_{\rm C}$) of a node is the reciprocal of the mean geodesic distance between the node and all other nodes reachable from it \cite{newman2012}. Therefore, it represents how fast information is spread from the node to other nodes in the network. Thus, for a node $i$, 
\begin{equation}
C_{\rm C}\ = \frac{n}{\sum_{j} {d_{ij}} },
\label{method_eq3}
\end{equation}  
where $d_{ij}$ represents the geodesic path length from nodes $i$ to $j$, and $n$ is the total number of vertices in the graph reachable from node $i$.      

\vskip 0.3cm
{\noindent}\textbf{\it 5. Betweenness centrality}\\
{\noindent}Betweenness centrality of a node is the measure of the extent to which the node has control over the communication of other nodes. Betweenness centrality ($C_{\rm B}$) of a node $v$ is computed as follows \cite{newman2005,brandes2001,mason2007}:
\begin{equation}
C_{\rm B}(i) = \sum_{s\neq i\neq t\in {\mathbf N}}\frac{\sigma_{st}(i)}{\sigma_{st}},
\label{method_eq4}
\end{equation} 
where $\mathbf N$ is the set of nodes, $s$ and $t$ are nodes in the graph different from $i$, $\sigma_{st}$ is the number of shortest path from $s$ to $t$, and path through $i$ in the case of $\sigma_{st}(i)$. The betweenness centrality value is normalized by dividing with the number of node pairs (excluding node $i$).

\vskip 0.3cm
{\noindent}\textbf{\it 6. Eigenvector centrality}\\
{\noindent}Eigenvector centrality of a node $i$ ($v_i$) in a network is proportional to the sum of $i$'s neighbor centralities \cite{bonacich1987}, and it is given by
\begin{eqnarray}
v_i=\frac{1}{\lambda}\sum_{j=nn(i)}v_j,
\end{eqnarray}
where $nn(i)$ indicates nearest neighbors of node $i$ in the network. $\lambda$ is the eigenvalue of the eigenvector $v_i$ given by
\begin{eqnarray}
Av_i=\lambda v_i,
\end{eqnarray}
where $A$ is the adjacency matrix of the network. The principal eigenvector of matrix $A$, which corresponds to maximum eigenvalue $\lambda_{\rm max}$, is taken to have positive eigenvector centrality scores \cite{canr2}.

\vskip 0.3cm
{\noindent}\textbf{\it 7. Modularity}\\
{\noindent}Finally, modularity is the measure of how well a network is divided in communities \cite{newman2004}. Modularity ($Q$) is express as follows:
\begin{equation}
Q = \frac{1}{2m}\sum_{ij}\left(A_{ij} - \frac{k_ik_j}{2m}\right)\delta(C_i,C_j),
\label{method_eq6}
\end{equation}
where $m$ is the total number of edges in the community, $A_{ij}$ is the adjacency matrix of size $i\times j$, $k$ represents degrees, and the $\delta$ function yields 1 if nodes $i$ and $j$ are in the same community.

\vskip 0.3cm
{\noindent}\textbf{\it Scaling nature of topological parameters}\\
{\noindent}The data of topological parameters (probability of degree distribution, clustering co-efficient, and neighborhood connectivity) and centrality parameters (betweenness, closeness, and eigenvector centralities) of the network, modules, and sub-modules at various levels (Fig. 2 and Fig. 6) in log-log plot show approximately parallel power-law fit lines. We follow one parameter scaling theory \cite{abra,pica,mack} to scale the data given by
\begin{eqnarray}
\frac{\Lambda}{k^\phi}=F\left[\frac{\xi}{k^{\phi}}\right],
\end{eqnarray}
where $F$ is a scaling function. For topological parameters $\Lambda(k)=P(k),C(k),c_n(k)$ and for centralities $\Lambda(k)=C_{\rm B}(k),C_{\rm C}(k),C_{\rm E}(k)$ with corresponding $\phi$ values after fit. The calculated $\xi$ after fitting each data of network/module/sub-module corresponds to the minimum path length of the network/module/sub-module approximately. This fitting procedure gives us $F\approx constant$. Hence, we found the following scaling law:
\begin{eqnarray}
\Lambda(k)\approx {\rm constant}\times k^\phi,
\end{eqnarray}
where $\phi=\{-\alpha,-\beta,-\gamma\}$ for $C(k)$, $C_n(k)$, and $P(k)$, respectively, and $\phi=\{\epsilon,\eta,\delta\}$ for $C_{\rm B}(k)$, $C_{\rm C}(k)$, and $C_{\rm E}(k)$, respectively.

\vspace{0.5cm}
\noindent {\bf Acknowledgments} \\
RKBS acknowledges CSIR, India, for providing financial support under sanction no. 25(0221)/13/EMR-II.

\end{document}